# Single-Molecule Vibrational Characterization of Binding Geometry Effects on Isocyanide-Metal Interactions

Liya Bi[1, 2 #], Zhe Wang[3, 4 #], Krista Balto[1], Andrea R. Tao[1, 2, 5], Tod A. Pascal[2, 5], Yanning Zhang[4], Joshua S. Figueroa[1, 2*] and Shaowei Li[1, 2*]

[1] Department of Chemistry and Biochemistry, University of California, San Diego, La Jolla, CA 92093-0309, USA

[2] Program in Materials Science and Engineering, University of California, San Diego, La Jolla, CA 92093-0418, USA

[3] Department of Physics and Astronomy, University of California, Irvine, CA 92697-4575, USA

[4] Institute of Fundamental and Frontier Sciences, University of Electronic Science and Technology of China, Chengdu, 611731, China

[5] Aiiso Yufeng Li Family Department of Chemical and Nano Engineering, University of California, San Diego, La Jolla, CA 92093-0448, USA

*Corresponding Author: jsfig@ucsd.edu (J.S.F.); shaoweili@ucsd.edu (S.L.)
#These authors contributed equally

**Abstract**

Isocyanide-metal binding is governed by σ-donation and π-back-bonding, which affects the energy of the isocyanide stretching mode—a characteristic probe for ligand-metal interactions. While extensive correlations exist between structure and spectroscopy in molecular isocyanide-metal systems, isocyanide interactions with metallic crystalline surfaces, where ligands often bind in various geometries, remain poorly understood. Conventional vibrational spectroscopies, such as infrared and Raman, lack the molecular-scale resolution needed to distinguish these inhomogeneous configurations. In contrast, inelastic electron tunneling spectroscopy with scanning tunneling microscopy (STM-IETS) enables direct visualization of ligand adsorption geometries and their vibrational signatures. Using STM-IETS, here we investigate an metal-adsorbed *m*-terphenyl isocyanide ligand and find that adsorption geometry on Cu(100) induces a significant shift in isocyanide stretching frequency, even greater than replacing Cu(100) with Ag(111). Density functional theory confirms this shift arises from atomic-scale variations in isocyanide-metal binding. This study elucidates how atomic-scale binding influences the vibrational signatures of isocyanide ligands—an often-overlooked factor in understanding isocyanide-metal interactions.

## Introduction

Molecular isocyanides (C≡NR), when functioning as ligands, binds to metals primarily through σ-donation and π-back-bonding.[1-3] In this interaction, the carbon atom of the isocyanide donates electron density from its filled, lone-pair-type orbital to the d orbital of the metal, forming a σ bond. Simultaneously, the metal returns electron from its filled d orbitals into the isocyanide vacant π* orbitals. This dual bonding mechanism modulates the electronic properties of both ligand and metal, influencing the stability and reactivity of isocyanide-metal complexes[4-6] – a core concept in coordination chemistry akin to the binding of carbon monoxide (CO).[7,8] Understanding these bonding modes is crucial for the design and application of isocyanide-metal complexes, especially in catalysis, where precise control of the metal's electronic properties can determine the efficiency and selectivity of reactions.[5,9,10]

The isocyanide stretching frequency, which often undergoes significant shift due to a changed electron density within the isocyanide bond, is a sensitive probe for this bonding interplay.[1,4,11] Generally, π-back-donation weakens the C≡N bond, lowering the stretching frequency, while σ-donation relieves C-N σ* character, strengthens the C≡N bond and shifts the stretching higher.[2,12-14] The balance between these opposing effects reflects the electron density and the coordination environment of the ligand-binding site, offering key insight into isocyanide-metal interactions.

On the surfaces of metallic nanoclusters or bulk crystals, various heterogeneous and complex adsorption geometries often coexist.[15-17] Even the subtle geometry difference in ligand's attachment—whether it binds in a linear or bent fashion,[17] or the number and type of metal atoms that the isocyanide binds to[18]—can affects how effectively electron density is transferred through both σ donation and π back-donation processes, crucially determining the strength and directionality of the ligand-metal interaction and ultimately altering the isocyanide stretching frequency and the local reactivity of the coordination site. However, conventional vibrational spectroscopies, such as infrared or Raman spectroscopy, lack the spatial resolution to distinguish these inhomogeneous geometries, limiting their ability to capture the atomic-scale structural details critical in isocyanide-metal interactions.[19,20]

To address this gap, we characterize the vibrational fingerprints of an *m*-terphenyl isocyanide ligand on metal surfaces using scanning tunneling microscopy (STM)[21,22] and inelastic electron tunneling spectroscopy (STM-IETS).[23,24] STM clearly resolves individual ligand binding geometries, while STM-IETS probes their spatially inhomogeneous vibrational properties.[16,25] Specifically, we investigate a rationally designed *m*-terphenyl ligand with a unique steric profile, CNAr$^{\text{Mes2}}$ (Ar$^{\text{Mes2}}$ = 2,6-(2,4,6-Me$_3$C$_6$H$_2$)$_2$C$_6$H$_3$)[6,26] (**Figure 1a**) on Cu(100) and Ag(111) surfaces. STM imaging reveals two distinct adsorption geometries on Cu(100): one straddling the Cu step edge and the other atop an adatom. Interestingly, STM-IETS reveals an ~10 meV (~80 cm$^{-1}$) shift in the Cu-bound isocyanide stretching frequency, $v$(Cu-C ≡ N), between these two geometries—a difference larger than that observed between molecules on Cu(100) and Ag(111) in a similar configuration. Density functional theory (DFT) captures this shift to be due to the atomic-scale variations in isocyanide-metal binding. These findings provide molecular-level insights into how adsorption geometry influences σ-donation and π-back-bonding, refining our understanding of isocyanide-metal interactions and their impact on chemical properties.

**Results and Discussion**

**Microscopic Visualization of Inhomogeneous Adsorption of $CNAr^{Mes2}$ on Cu(100) Surface**

Individual $CNAr^{Mes2}$ ligands on Cu(100) surface exhibit two binding geometries, clearly distinguishable in STM topographies. The molecules were deposited onto surfaces at 5.2 K, then gradually warmed to room temperature to reach thermodynamically favored adsorption sites. The sample was then cooled back to 5.2 K for STM characterization (see details in the Experimental Methods section of Supporting Information). When $CNAr^{Mes2}$ binds to the metal surfaces, steric pressure from the bulky aryl groups disrupts the direct isocyanide-metal interaction on planar domains, instead promoting selective adsorption at high-curvature sites or adatoms.[25, 27, 28] Our previous studies on Au(111) and Ag(111) revealed two distinct mechanisms that relieve this steric strain: $CNAr^{Mes2}$ either straddles step edge of Au where surface curvature is high,[25] or binds atop an Ag adatom and resides parallelly to the step edge.[27] Interestingly, on Cu(100) surface, both adsorption mechanisms co-exist. STM image (**Figure 1b**) shows two distinct appearances of the adsorbed molecules on Cu(100). The zoom-in image to the step edges reveals a mixture of two adsorption geometries. One species appears as an elliptical protrusion rotated ~50° from the step edge, resembling the "straddling" geometry observed on Au(111) (**Figure 1c**).[25] The other species shows two asymmetric lobes aligned parallel to the step edge, consistent with the "parallel" adatom-bound geometry seen on Ag(111) where the smaller lobe is attributed to isocyanide adsorption onto a surface adatom (**Figure 1d**).[27] These two geometries were previously identified through high-resolution STM images with a molecularly functionalized tip on Au and Ag. Though high-resolution structural imaging on Cu(100) is limited by challenges in tip functionalization, the resemblance to known configurations on Au and Ag supports the same underlying binding mechanisms. In subsequent sections, we refer to these two configurations as "straddling" and "parallel," corresponding to the step-edge and adatom-binding geometries, respectively.

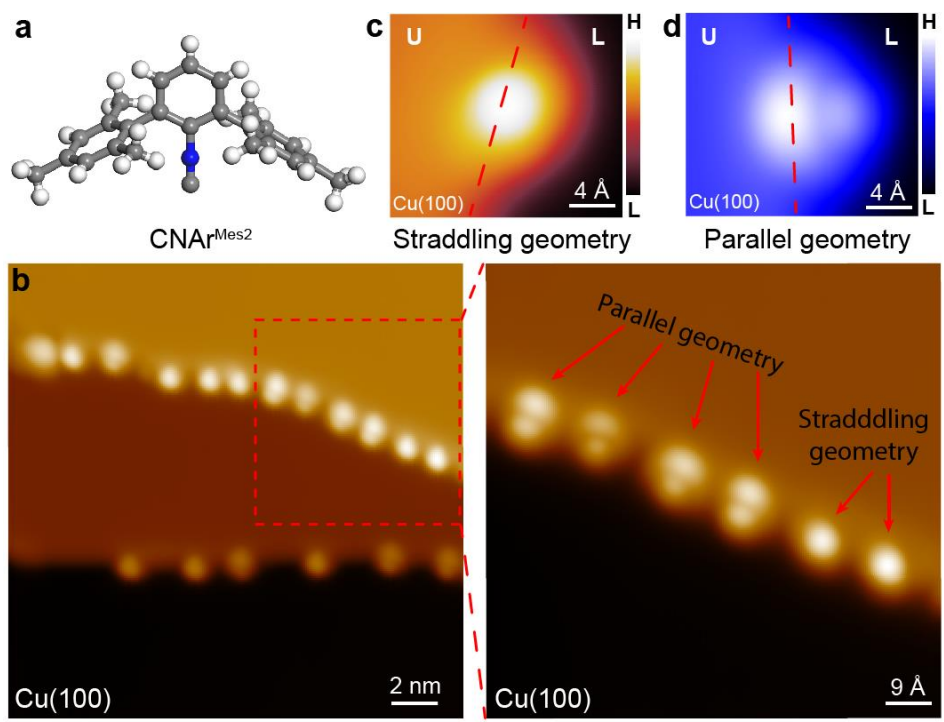

**Figure 1.** Adsorptions of CNAr$^{Mes2}$ on Cu(100) surface. (**a**) Molecular structure of CNAr$^{Mes2}$. C, H and N atoms are shown in grey, white and blue, respectively. (**b**) STM image of CNAr$^{Mes2}$–bound Cu(100) surface. (**c, d**) Zoom-in images of single CNAr$^{Mes2}$ in straddling (c) and parallel geometry (d). The red dashed lines indicate the step edges. Upper (U) and lower (L) atomic layers are labelled for clarity. Imaging parameters were 1.5 V, 20 pA.

## Vibrational Analysis of CNAr$^{Mes2}$ on Cu(100) Surface

To investigate how different adsorption geometries influence the isocyanide-metal interaction, we used STM-IETS to probe the vibrations of CNAr$^{Mes2}$ in both straddling and parallel configurations. As shown in **Figure 2a**, both geometries exhibit seven vibrational modes (labeled I-VII) below 150 meV (~1210 cm$^{-1}$). While signal intensities differ—likely due to variations in vibrational excitation cross-sections[29] or site-dependent spatial confinement[25, 30]—the mode energies remain largely consistent, with no discernible variation in energies of both low-energy modes (**Figure 2a**) and the ring breathing mode at ~195 meV (**Figure 2b**) between two geometries. In contrast, the $\nu$(Cu-C≡N) vibration exhibits a pronounced ~80 cm$^{-1}$ shift (**Figure 2b**), highlighting a significantly different bonding interaction between the two configurations.

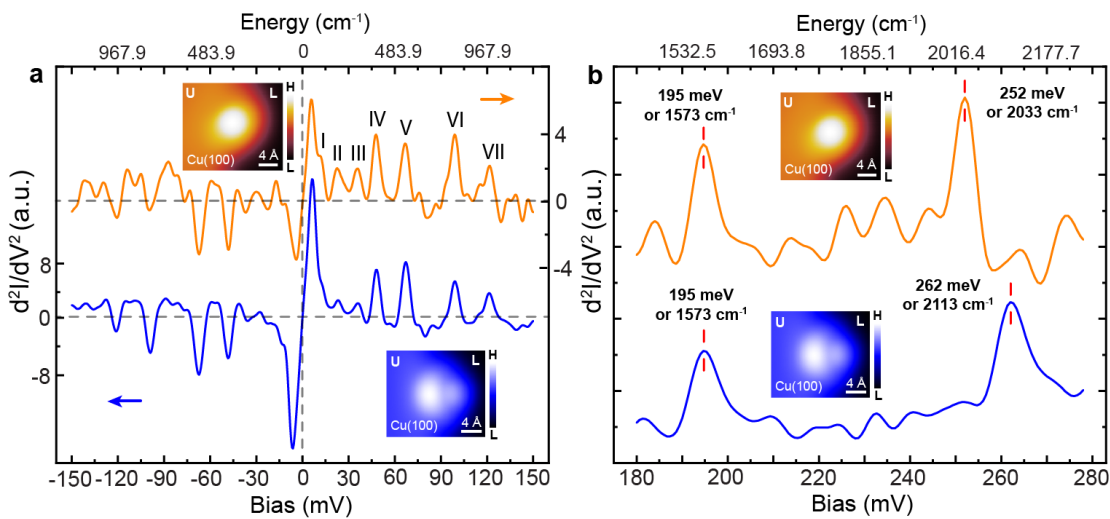

**Figure 2.** Vibrational characterization of CNAr$^{Mes2}$ on Cu(100) surface. (**a, b**) STM-IETS of CNAr$^{Mes2}$ in straddling (orange spectra and top insets) and parallel (blue spectra and bottom insets) geometries that show a series of vibrational modes. The spectra are vertically shifted for clarity. Imaging parameters were 1.5 V, 20 pA. Spectroscopy parameters were 150 mV, 500 pA (a) and 300 mV, 2 nA (b).

To uncover the origins of the observed shift in $\nu$(Cu-C≡N) mode, we performed DFT-based vibrational density of states (vDOS) calculations. As shown in **Figure 3**, the simulated vDOS closely matches the experimental STM-IETS results (**Figure 2**), reproducing all seven low-energy modes (I–VII) below 150 meV (**Figure 3a**), which are primarily associated with collective motions of the *m*-terphenyl group (see the simulated nuclear motions in **Figure S2**). These modes show negligible energy differences between straddling and parallel geometries, consistent with the experiment and indicating minimal ligand structural change upon adsorption. Differences in signal intensity between simulation and STM-IETS likely reflect selection rule effects, which remains under development for STM-IETS.[31,][32] Importantly, DFT simulation captures the significant shift in $\nu$(Cu–C≡N) mode (**Figure 3b** and **Figure S3**) though the simulated energy difference (~15 meV) slightly exceeds the experimental value (~10 meV). Additionally, the vDOS suggests two ring-breathing vibrations around 195 meV for CNAr$^{Mes2}$ in both geometries (**Figure 3b**). However, their energy difference is minimal, making them indistinguishable in STM-IETS.

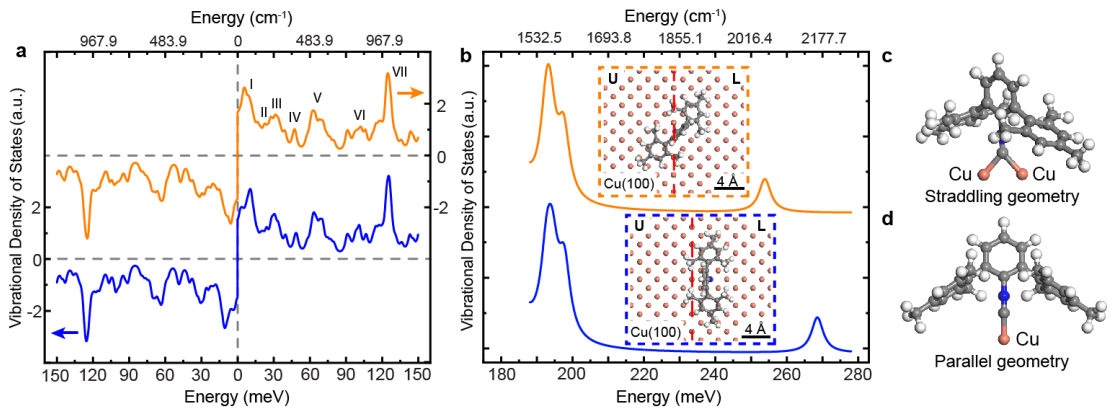

**Figure 3.** Calculated adsorptions and vibrations of CNAr[Mes2] on Cu(100) surface. (**a, b**) vDOS of CNAr[Mes2] in straddling (orange spectra) and parallel (blue spectra) geometry on Cu(100) surface. The spectra are vertically shifted for clarity. The top/bottom inset presents the top view of the relaxed adsorption of straddling/parallel CNAr[Mes2] on Cu(100) surface. The red dashed lines indicate the step edges. Upper (U) and lower (L) atomic layers are labelled for clarity. (**c, d**) Side view of CNAr[Mes2] in the straddling (c) and parallel (d) geometry, highlighting the bridging feature in the former and the end-on adsorption in the latter.

DFT simulations further reveal that the significant shift in $v(Cu–C≡N)$ between the two adsorption geometries arises from differences in atomic-scale binding configurations between isocyanide and Cu atoms. As shown in **Figure 3b-d** and **Figure S4**, both the molecules in the relaxed straddling and parallel geometries effectively relieve steric repulsion between the mesityl side groups of CNAr[Mes2] and the surface, resulting in similar adsorption energies. A key distinction lies in the number of surface Cu atoms binds with the isocyanide: in the straddling geometry, the isocyanide bridges two Cu atoms at a step-edge trough site (**Figure 3c**), while in the parallel geometry, it binds to a single Cu adatom (**Figure 3d**). Despite comparable steric environments—confirmed by similar bouncing (~48 meV, mode IV) and methyl-twisting (~67 meV, mode V) modes, both of which are sensitive to ligand-surface steric interactions,[25] the $v(Cu–C≡N)$ mode remains highly sensitive to this difference in local coordination. DFT calculations further confirm the critical role of this local binding geometry between the isocyanide and Cu atoms in determining the energy of $v(Cu-C≡N)$. For example, when the molecule binds directly to the atop site of a Cu step edge, compared to a trough site, the simulated $v(Cu–C≡N)$ exhibits a blue shift of ~14 meV, while other vibrational energies remain unchanged (**Figure S5**).

The frequency shift of the metal-bound isocyanide stretching mode reflects the balance between σ-donation and the π-back-bonding.[1, 2, 3, 12] The $v(Cu-C≡N)$ of CNAr[Mes2] in the parallel geometry (**Figure 2b**, blue) is close to that of unbound molecule in solution,[26] suggesting that the σ-donation is counterbalanced by the weak π-back-donation. This spectroscopic signature is in agreement with previous studies on mononuclear *m*-terphenyl isocyanide Cu(I) complexes, in which σ-donation to the metal from terminal isocyanide ligands is the dominant bonding interaction and π-back-bonding is largely absent due to contracted d-orbital manifold of the $d^{10}$ Cu(I) ion.[26, 33-35] In contrast, when CNAr[Mes2] adsorbs in the straddling geometry where the isocyanide group bridges two adjacent Cu atoms, the significantly redshifted $\nu$(Cu-C≡N) band (**Figure 2b**, orange) is indicative of enhanced π-back-donation to the isocyanide π* orbitals.[4, 9, 36] As has been well established in molecular systems for bridging isocyanides, as well as the related CO ligand, electronic population of π* orbitals from π-symmetry d-orbitals from two adjacent metal centers is a central component of the bridging coordination mode.[37] While three-center-two-electron (3c-2e⁻) σ-donation is also present in the bridging coordination mode, the direct overlap of the isocyanide π* orbitals with filled π-symmetry d-orbitals from two adjacent Cu atoms results in significant bond weakening and red-shifting of the $\nu$(Cu-C≡N) band in a manner not compensated by σ-donation.[36, 37] This π-symmetry orbital overlap is weaker when the isocyanide directly binds in a terminal fashion to a Cu adatom, leading to enhanced σ-donation, but attenuated π-back-donation.[12-14, 38] As a result, this difference in atomic-scale ligand-surface interactions gives rise to the observed variation in $\nu$(Cu-C≡N) between two adsorption geometries. It is worth noting that the red-shift of isocyanide stretching frequency is commonly reported in isocyanide ligands binding to various metal nanocrystalline surfaces[15, 18], where we believe a similar binding mode—where the isocyanide interacts with multiple surface atoms—also plays a critical role.

**Vibrational Comparison between the Binding of CNAr[Mes2] to Ag(111) and Cu(100) Surfaces in a Similar Geometry**

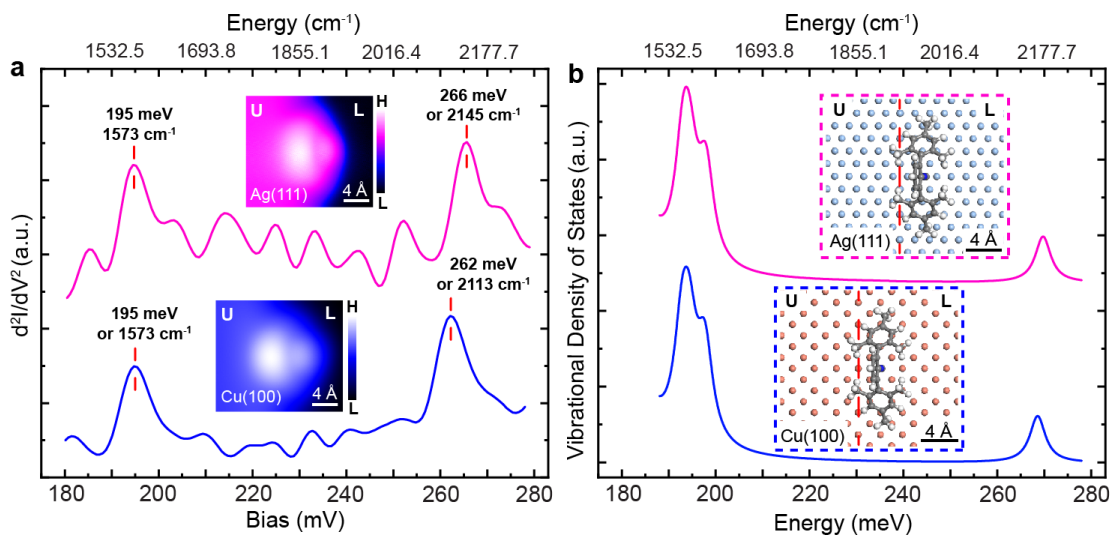

**Figure 4.** Vibrational comparison between CNAr[Mes2] on Ag(111) and Cu(100) surfaces. (**a, b**) STM-IETS (a) and vDOS (b) of CNAr[Mes2] on Ag(111) (magenta) and Cu(100) (blue) surfaces in a parallel geometry. The spectra are vertically shifted for clarity. Imaging and spectroscopy parameters in (a) were -1 V, 20 pA (top inset) and -300 mV, 1 nA (magenta spectrum); 1.5 V, 20 pA (bottom inset) and 300 mV, 2 nA (blue spectrum).

Because isocyanides can engage in both σ-donation and π-back-donation, their bonding is highly sensitive to the metal to which they bind, and is widely used as a spectroscopic probe for chemical identification.[18, 39-41] However, it remains a key question whether the dominant factor behind shifts in isocyanide stretching is the metal identity or the local binding geometry. To address this, we further perform STM-IETS on CNAr$^{Mes2}$ adsorbed on Ag(111), where, as previously reported, the molecule primarily adopts a parallel geometry—binding atop a single Ag adatom.[27] As shown in the insets of **Figure 4a**, the STM images of CNAr$^{Mes2}$ on Ag(111) closely resemble the parallel geometry on Cu(100), suggesting a similar binding geometry. STM-IETS measurements reveal $\nu$(Ag-C≡N) at ~266 meV (~2145 cm$^{-1}$), just ~4 meV (~32 cm$^{-1}$) higher than the measured $\nu$(Cu-C≡N) for the parallel species on Cu(100). This minor energy difference aligns with DFT simulations (**Figure 4b**), which predict a slightly higher isocyanide stretching energy on Ag(111) (also shown in **Figure S3** and **Figure S6**). However, this metal-dependent shift is significantly smaller than the ~10 meV energy difference observed between the two differing isocyanide geometries on Cu(100). Importantly, the energetic similarity of the $\nu$(M-C≡N) stretches for CNAr$^{Mes2}$ bound to either Cu(100) or Ag(111) are reflective of similar charge distributions of the surface metal atoms with respect to their ability to engage in σ-donation and π-back-donation from and to the isocyanide, respectively.[42] This is remarkably similar to molecular transition-metal carbonyl complexes (i.e. M(CO)$_n$), which show nearly invariant $\nu$(M-C≡O) stretches for metals in a periodic group,[42] irrespective of the radial extension and overlap of the 3d, 4d, or 5d orbitals. Therefore, for the coinage-metal surfaces studied here, these comparative spectroscopic responses, coupled with high-resolution imaging, can help identify ligand binding interactions as a function of geometry in a manner that is commonplace for molecular systems. Accordingly, these findings underscore the crucial role of atomic-scale binding structures in determining the chemical properties of ligand-metal interactions and highlight the importance of characterizing and understanding ligand adsorption geometries with molecular-scale structure resolution when interpreting vibrational energy shifts.

## Conclusions

In conclusion, our study shows that the vibrational properties of isocyanide-metal interactions are shaped not only by the metal identity but also, if not more critically, by the atomic-scale binding geometry. Using STM and STM-IETS, we demonstrate that the distinct adsorption geometries of CNAr$^{Mes2}$ on Cu(100) surface lead to notable differences in the isocyanide stretching frequency, an energy shift that is more significant than the effect from substituting the substrate with Ag(111) surface. The observed vibrational shifts within the same metal surface, Cu(100), highlight the dominant influence of the atomic-scale isocyanide binding geometry over elemental composition in determining σ-donation/π-back-donation strength. DFT simulations further confirm that variations in local coordination environments govern the extent of orbital overlap, ultimately shaping the electronic properties of the isocyanide-metal interaction. These findings underscore the necessity of precise molecular-scale structural characterization when interpreting vibrational spectroscopy results and provide deeper insights into the fundamental factors governing isocyanide-metal bonding. Understanding these relationships is crucial for the rational design of metal-ligand interfaces in catalysis,[43] molecular sensing,[44] and surface chemistry applications.[38]

## Acknowledgment

The authors acknowledge the use of facilities and instrumentation supported by the United States National Science Foundation (NSF) through the UC San Diego Materials Research Science and Engineering Center (UCSD MRSEC) with Grant No. DMR-2011924. This work was primarily supported by the United States Department of Energy (DOE) under Grant DE-SC0025537 (to Andrea Tao and Shaowei Li), and National Science Foundation (NSF) under Grant DMR-2011924 (UCSD MRSEC) and CHE-2303936 (to Shaowei Li). This work also used the Expanse supercomputer at the San Diego Supercomputing Center through allocation CSD799 from the Advanced Cyberinfrastructure Coordination Ecosystem: Services & Support (ACCESS) program, which is supported by NSF grants No. 2138259, No. 2138286, No. 2138307, No. 2137603, and No. 2138296.